\documentclass{eptcs}
\providecommand{\event}{ACL2-2018} 
\usepackage{breakurl}             
\usepackage{underscore}           

\usepackage{tikz}
\usepackage{amsmath}
\usepackage{amsthm}
\usepackage{amssymb}
\usepackage{enumitem}
\usepackage{mathtools}
\usepackage{thmtools}
\usepackage{enumitem}
\usepackage{nameref, hyperref, cleveref}
\usepackage{tikz}
\usepackage[title]{appendix}
\usepackage{float}
\usepackage{listings}

\usetikzlibrary{arrows.meta}

\DeclareMathOperator{\st}{st}

\crefname{equation}{}{}
\crefname{program}{Program}{Programs}

\providecommand{\thisvolume}[1]{this volume of EPTCS, Open Publishing Association}

\lstdefinestyle{progstyle}{
    aboveskip=3mm,
    belowskip=3mm,
    showstringspaces=false,
    columns=flexible,
    basicstyle={\small\ttfamily},
    numbersep=5pt,
    numberstyle=\tiny\color{gray},
    keywordstyle=\color{blue},
    commentstyle=\color{green},
    stringstyle=\color{purple},
    breaklines=true,
    breakatwhitespace=true,
    tabsize=8,
    keepspaces=true
  }

\floatstyle{ruled}
\newfloat{Program}{htbp}{lop}[section]

\usetikzlibrary{rulercompass}

\theoremstyle{definition}
\declaretheorem[name = Theorem,  refname = {Theorem, Theorems}]{thm}

\declaretheorem[name = Lemma, sibling = thm, refname = {Lemma}]{lem}

\title{Convex Functions in ACL2(r)}
\author{Carl Kwan \qquad Mark R. Greenstreet
\institute{Department of Computer Science \\
University of British Columbia\thanks{
This work is supported in part by the National Science 
and Engineering
Research Council of Canada (NSERC) Discovery Grant program and
the Institute for Computing, Information and Cognitive Systems 
(ICICS) at UBC.}
\\
Vancouver, Canada}
\email{\{carlkwan, mrg\}@cs.ubc.ca}
}
\def\titlerunning{Convex Functions in ACL2(r)}
\def\authorrunning{Carl Kwan \& Mark R. Greenstreet}

\newcounter{tempcounter}
\newcounter{ineqcounter}
\newenvironment{ineq}
 {%
  \setcounter{tempcounter}{\value{equation}}%
  \setcounter{equation}{\value{ineqcounter}}%
  \gather
 }
 {%
  \endgather
  \setcounter{equation}{\value{tempcounter}}%
  \setcounter{ineqcounter}{\value{equation}}%
  \stepcounter{ineqcounter}
 }

\newcommand{\Rn}[1]{\ensuremath{\mathbb{R}^{#1}}}
\newcommand{\convF}[1]{\ensuremath{\mathcal{F}^{#1}}}
\newcommand{\FL}{\ensuremath{\mathcal{F}^1_L}}
\newcommand{\FLn}[1]{\ensuremath{\FL(\Rn{#1})}}

\begin{document}
\maketitle



\begin{abstract}

This paper builds upon our prior formalisation of
$\Rn{n} $ in ACL2(r) by presenting a 
set of theorems for reasoning about
convex functions. This is a demonstration of the higher-dimensional
analytical reasoning possible in our metric space formalisation of $\mathbb
R^n$. Among the introduced theorems is a set of equivalent conditions for convex
functions with Lipschitz continuous gradients from Yurii Nesterov's classic
text on convex optimisation. To the best of our knowledge a full
proof of the theorem has yet to be published in a single piece of literature.
We also explore ``proof engineering'' issues, such as how to state 
Nesterov's theorem in a manner that is both clear and useful.

\end{abstract}

\section{Introduction}

Convex optimisation is a branch of applied mathematics that finds widespread
use in financial modelling, operations research, machine learning, and many
other fields.
Algorithms for convex optimisation often have many parameters that can be
tuned to improve performance.  However, a choice of parameter values that
produces good performance on a set of test cases may suffer from poor
convergence or non-convergence in other cases.  Hand written proofs for
convergence properties often include simplifying assumptions to make the
reasoning tractable.  This motivates using machine generated and/or verified
proofs for the convergence and performance of these algorithms.  Once an
initial proof has been completed, the hope is that simplifying assumptions
can be incrementally relaxed or removed to justify progressively more aggressive
implementations.
These observations motivate our exploration of convex functions within ACL2(r).

We present example proofs of continuity,  Lipschitz continuity, 
and convexity for some simple functions as well as some basic theorems 
of convex optimisation.
To the best of our knowledge, there are no other formalisations of convex
functions in published literature 
(though we were able to find some formal 
theorems
involving convex hulls in \cite{harrison2007}).
Moreover, we also provide a proof for a set of equivalent conditions for 
inclusion in the class of convex functions with Lipschitz continuous gradients --
a theorem that, 
to the best of our knowledge,
has yet to be fully published in a single piece of literature with a correct proof.
This characterisation is based on Yurii Nesterov's 
classic work on convex optimisation~\cite{Nesterov2004} which has 
applications in the convergence proofs for many gradient descent algorithms.

This paper builds on the formalisation of $\mathbb R^n$ 
as an inner product space and as a metric space in~\cite{Kwan2018-cs}.
Of particular note, we make use of the Cauchy-Schwarz inequality 
which was a demonstration of the algebraic reasoning capable in 
such a formalisation.
This subsequent paper demonstrates the analytical reasoning about multivariate functions 
$\Rn{n}\to \mathbb R$ that is enabled by the theorems proven in the books from the previous paper.
In addition to presenting some key lemmas, we also discuss some of the 
challenges of formalising theorems with
proofs that rely heavily on informally well-established and intuitive notions.

\section{Preliminaries}
This section summarizes the formalisation of vector spaces, inner-product spaces,
metric spaces, and the Cauchy-Schwarz inequality that are presented in 
our previous paper~\cite{Kwan2018-cs}.
The ACL2(r) formalisation is provided in the ACL2 books that
accompany these papers.
This work depends greatly on the 
original formalisation of the reals via non-standard analysis  in ACL2(r)~\cite{Gamboa2001}.
Further background on non-standard analysis can be found in \cite{robinson,cutland, loeb}.
 Background on vector, inner product, and metric spaces 
can be found in~\cite{shilov, roman, babyrudin, lang, jacob}.
We also outline some theorems involving convex functions.
Standard texts on convex optimisation include \cite{boyd, Nesterov2004}.

\subsection{Inner Product \& Metric Spaces}

An inner product space is a vector space $(V,F)$ equipped with an inner product 
$\langle -, -\rangle:V\to F$. The inner product satisfies 
\begin{align}
&a\langle u, v\rangle = \langle a u ,v\rangle\label{eq:inner-prod.scalar-mult} \\
&\langle u + v , w \rangle = \langle u, w\rangle + \langle v, w\rangle\label{eq:inner-prod.dist} \\
&\langle u,v\rangle = \langle v, u\rangle  \text{ when } F=\mathbb R\label{eq:inner-prod.commute} \\
& \langle u,u\rangle \geq 0\text{ with equality iff }u = 0\label{eq:inner-prod.uu-non-neg}
\end{align}
for any $u,v,w\in V$ and $a\in F$~\cite[Chapter 9]{roman}.

An inner product induces a norm $\|\cdot \|$. If $\|\cdot\| = \sqrt{\langle -,-\rangle}$, 
then the Cauchy-Schwarz inequality~\cite[Chapter 15]{lang} holds for any $x,y\in V$:
\begin{equation}\label{eq:cauchy-schwarz}
|\langle x , y\rangle |\leq \|x\|\|y\|.
\end{equation}

Norms induce metrics~\cite[Chapter 2]{babyrudin} $d(x,y) = \|x-y\|$ which satisfy 
\begin{align}
d(x,y) = 0 \iff x= y && \textit{(definitness)}\label{eq:dist.eq0} \\
d(x,y)=d(y,x)  && \textit{(symmetry)}\label{eq:dist.symmetric}\\
d(x,y) \leq d(x,z) + d(z, y) &&\textit{(triangle inequality)}.\label{eq:dist.triangle}
\end{align}
From this it follows that 
\begin{equation}
d(x,y)\geq 0\label{eq:dist.non-neg}
\end{equation}
because 
\begin{displaymath}
0=d(x,x) \leq d(x,y) + d(y, x) = 2d(x,y).
\end{displaymath}
A metric space is a pair $(M, d)$ where $M$ is a set and $d$ is a metric on $M$~\cite[Chapter 2]{babyrudin}.

A function $f:M\to M^\prime$ between metric spaces is continuous~\cite[Chapter 4]{babyrudin} if for every $x\in M$ and
for any $\epsilon >0$, there is a $\delta>0$ such that 
\begin{equation}\label{eq:continuity}
d_M(x,y) < \delta \implies d_{M^\prime}(f(x), f(y)) <\epsilon
\end{equation}
for any $y\in E$.

Throughout this paper, we will consider $\Rn{n}$ adjoined with 
the dot product and Euclidean metric to be an inner product and metric space, respectively.

\subsection{Continuity \& Differentiability}
A univariate function $f:\mathbb R\to\mathbb R$ is continuous
if
for any $x\in\mathbb R$ and
 $\epsilon>0$, there is a $\delta>0$ such that for any 
$y\in\mathbb R$,
if $|x-y|<\delta$, then $|f(x)-f(y)|<\epsilon$~\cite[Chapter 4]{babyrudin}. 
Moreover, the derivative of 
$f$ is defined to be
\begin{equation}\label{eq:deriv.def}
f^\prime (x) = \lim_{h\to0} \frac{f(x+h)-f(x)}h
\end{equation}
if such a form exists~\cite[Chapter 5]{babyrudin}.

For a multivariate function $f:\Rn{n} \to \mathbb R$, continuity is 
defined similarly.
We call $f$ continuous if 
for any $x\in\mathbb R^n$ and $\epsilon>0$, 
there is a $\delta>0$ such that for any $y\in\mathbb R^n$, 
if $\|x-y\|<\delta$, then 
$|f(x)-f(y)|<\epsilon$~\cite[Chapter 4]{babyrudin}.
Likewise, if it exists, 
there is a derivative for multivariate functions defined similarly to the 
univariate case:
\begin{equation}\label{eq:deriv-multivar.def}
\langle f^\prime(x), h\rangle = \lim_{\|h\|_2\to 0} \frac{f(x+h) - f(x)}{\|h\|_2}.
\end{equation}
We call $f^\prime:\Rn{n}\to\Rn{n}$ the \textit{gradient} of  $f$ and it satisfies 
\begin{equation}
f^\prime(x) = (f^\prime_1(x_1), f^\prime_2(x_2), \hdots, f^\prime_n(x_n))
\end{equation}
where $f^\prime_i(x_i)$ is the univariate derivative of $f$ with respect to the $i$-th 
component $x_i$ of $x$~\cite[Chapter 9]{babyrudin}.

\subsection{Convex Functions \& $\FLn{n}$}

A function $f:\mathbb R^n \to \mathbb R$ 
is convex~\cite[Chapter 3]{boyd} if for any $x,y\in\Rn{n}$ and $\alpha\in[0,1]$,
\begin{equation}\label{eq:convex-fn.def}
  \alpha f(x) + (1-\alpha)f(y) \geq f(\alpha x+(1-\alpha)y). 
\end{equation}
Equivalently~\cite[Def.\ 2.1.1]{Nesterov2004}, if $f$ is differentiable once with gradient $f^\prime$, then 
it is convex if 
\begin{equation}
\label{eq:convex-fn-nest.def}
f(y) \geq f(x) + \langle f^\prime (x), y-x\rangle.
\end{equation}
Following Nesterov, we write $\mathcal F(\mathbb R^{n})$
 to denote the class of convex functions from $\Rn{n}$
to $\mathbb{R}$.
Examples of convex functions include $f(x)=x^2$, $\|\cdot\|_2$, and $\|\cdot\|_2^2$.
Moreover, the class of convex functions is closed under certain operations~\cite[Chapter 3]{boyd}.
\begin{thm}\label{thm:compose}
If 
$f:\mathbb R^n\to \mathbb R$ and
 $g:\mathbb R^n\to\mathbb R$ and
 $h:\mathbb R\to\mathbb R$ are convex with
$h$ monotonically increasing, then
\begin{enumerate}
    \item $a\cdot f$ is convex for any real $a\geq 0$,
    \item $f+g$ is convex,
    \item $h\circ f$ is convex.
\end{enumerate}
\end{thm}
Informal proofs of these claims follow from the definitions and can be found in 
\cite[Chapter 3]{boyd}.

Often, convex optimisation algorithms require $f$ to be both convex and
sufficiently ``smooth". Here, we take ``smooth" to be stronger than continuous
but not necessarily differentiable.
In particular,
we say that $f: \Rn{n} \rightarrow \mathbb{R}$ is Lipschitz continuous
if for any $x,y\in\Rn{n}$ there is some $L>0$ such that
\begin{equation}\label{eq:lipschitz.def}
\|f(x) - f(y)\| \leq L \|x - y\|.
\end{equation}
Informally, we have the following chain of inclusions for 
 classes of functions:
\[
\text{Differentiable } \subset \text{ Lipschitz Continuous } \subset \text
{ Continuous} .
\]

We write $\FLn{n}$ for the class of convex differentiable functions on
$\Rn{n}$ with Lipschitz continuous gradient with constant $L$.
Functions in the class $\FLn{n}$ have many useful properties for optimisation.
The main result of this paper is proving a theorem from~\cite[Thm.\ 2.1.5]{Nesterov2004}
that gives six ``equivalent'' ways of showing that a convex function is in $\FLn{n}$:
\begin{thm}[Nesterov]\label{thm:nesterov}
Let $f \in \mathcal F^1(\mathbb R^n)$, 
$x,y\in\Rn{n}$ and $\alpha\in[0,1]$. The following conditions are 
equivalent to $f\in\mathcal F_L^1(\mathbb R^n)$:
\begin{ineq}
  f(y)  \leq f(x) + \langle f^\prime(x) , y - x\rangle + \frac L 2\|x-y\|^2\label{nest:1}\\
  f(x) + \langle f^\prime(x) , y - x\rangle + \frac 1 {2L}\|f^\prime(x)-f^\prime(y)\|^2  \leq f(y)\label{nest:2}\\
 \frac 1L\|f^\prime(x) - f^\prime(y)\|^2 \leq \langle f^\prime (x) - f^\prime(y), x -y\rangle\label{nest:3}\\
\langle f^\prime(x) - f^\prime(y), x- y\rangle \leq L \|x-y\|^2\label{nest:4}\\
f(\alpha x+(1-\alpha)y)+\frac{\alpha(1-\alpha)}{2L}\|f^\prime(x) -f^\prime(y)\|^2
\leq \alpha f(x) +(1-\alpha)f(y)\label{nest:5}\\
\alpha f(x) +(1-\alpha)f(y) \leq
f(\alpha x+(1-\alpha)y)+\alpha(1-\alpha)\frac{L}{2}\|x-y\|^2.\label{nest:6}
\end{ineq}
\end{thm}
There are several motivations to prove Thm.~\ref{thm:nesterov} in ACL2(r).
First, such a formalisation provides an unambiguous statement of the theorem.
For example, the theorem requires the assumption $f \in \convF{n}$, but this
hypothesis is not explicitly stated in the theorem statement in~\cite{Nesterov2004}.
Instead, the assumption is implicit in the preceding text.
On the other hand, \ref{nest:5} implies Eq.~\ref{eq:convex-fn.def}
and therefore that $f$ is convex.
Inequalities~\ref{nest:2} through~\ref{nest:6} implicitly have an existential quantification of $L$.
By stating and proving the theorem in ACL2(r), these ambiguities are avoided.
Furthermore, this enables the use of 
Nesterov's theorem for further reasoning about
convex functions and optimisation algorithms.

\subsection{ACL2(r) \& Non-standard Analysis}

The usual axioms for $\Rn{n}$ as an inner product  and metric space were formalised in
ACL2(r) in \cite{Kwan2018-cs} along with theorems proving their salient properties.
Reals and infinitesimals are recognized by \texttt{realp} and \texttt{i-small}, respectively.
Two reals are \texttt{i-close} if their difference is \texttt{i-small}.
Vectors are recognized by \texttt{real-listp}.
Vector addition and subtraction are \texttt{(vec-+ x y)} and \texttt{(vec-- x y)}, respectively.
 Scalar multiplication is 
\texttt{(scalar-* a x)}.
The dot product is simply \texttt{(dot x y)}.
 The Euclidean norm and metric are \texttt{(eu-norm x)}
and \texttt{(eu-metric x y)}, respectively. 
Sometimes it is easier to reason about the 
square of the norm or metric. These are called 
\texttt{(norm\textasciicircum2 x)} and \texttt{(metric\textasciicircum2 x y)}, respectively. 
More details can be found in~\cite{Kwan2018-cs}.

In non-standard analysis, continuity for a function $f:\Rn{n} \to \mathbb R$ 
amounts to
if  $\|x-y\|$ is an infinitesimal for some standard $x$, 
then so is $|f(x) - f(y)|$.
The derivative of a function $f:\mathbb R\to \mathbb R$ is the standard part of 
$
 \frac{f(x+h)-f(x)}h
$
where $h$ is an infinitesimal.
The formalisation for continuity, differentiability, integrability, and 
the Fundamental Theorem of Calculus already exist for univariate 
functions in ACL2(r).
\cite{Gamboa2000, Kaufmann2000}

\section{Convexity in ACL2(r)}
In this section, we provide some selected examples of formalised theorems involving 
convex functions. 
The formalised proofs follow almost directly from those of the informal proofs.
For the sake of exposition, the proof for the first theorem 
is outlined but the rest are omitted. 

The first is a simple theorem positing the convexity of 
$f(x)=x^2$ which is true because for $0 \leq \alpha \leq 1$,
\begin{equation}\label{eq:square}
\alpha x^2 + (1-\alpha)y^2 - (\alpha x + (1-\alpha)y)^2 
= 
\alpha(1-\alpha)(x^2 - 2xy + y^2)
=
\alpha(1-\alpha)(x-y)^2 \geq 0.
\end{equation}
We first define a square function:
\texttt{(defun square-fn (x) (* (realfix x) (realfix x)))}.
The chain of equalities in Eq.~\ref{eq:square} is immediately recognized by ACL2(r).
The inequality in Eq.~\ref{eq:square} also passes without issue.  
The convexity of \texttt{square-fn} then follows from a simple application of the two lemmas.

\begin{Program}\label{prog:square}
    \caption{The equality in Eq.~\ref{eq:square} formalised.}
    \begin{lstlisting}
;; ax^2 + (1-a)y^2 - (ax + (1-a)y)^2 = a(1-a)(x-y)^2
(defthm lemma-1
 (implies (and (realp x) (realp y) (realp a) (<= 0 a) (<= a 1))
          (equal (- (+ (* a (square-fn x)) (* (- 1 a) (square-fn y)))
                    (square-fn (+ (* a x) (* (- 1 a) y))))
                 (* a (- 1 a) (square-fn (- x y))))))
\end{lstlisting}
\end{Program}
\begin{Program}\label{prog:square-gen}
    \caption{A more general version 
        of the inequality in Eq.~\ref{eq:square} formalised.}

        \begin{lstlisting}
;; replace a with a(1-a) and x with x-y to obtain the desired inequality
(defthm lemma-2
 (implies (and (realp a) (<= 0 a))
          (<= 0 (* a (square-fn x)))))
\end{lstlisting}
\end{Program}
\begin{Program}
    \caption{The square function is convex.}
    \begin{lstlisting}
(defthm square-fn-is-convex
 (implies (and (realp x) (realp y) (realp a) (<= 0 a) (<= a 1))
          (<= (square-fn (+ (* a x) (* (- 1 a) y)))
              (+ (* a (square-fn x)) (* (- 1 a) (square-fn y)))))
 :hints (("GOAL" :use ((:instance lemma-2 (a (* a (- 1 a))) (x (- x y)))))))
\end{lstlisting}
\end{Program}


Also formalised is a proof of each of the statements in Thm.~\ref{thm:compose}. 
Here we outline the proof of the convexity of $a\cdot f$ given that $a\geq 0$ and 
$f$ is convex.
The rest are similar. Moreover, the approach we 
take resembles our approach to formalising Thm.~\ref{thm:nesterov} albeit much simpler. 
In particular, to reason about functions, we use the technique of encapsulation 
to  first prove the desired theorem for a witness function.
Functional instantiation then provides a method for
reasoning about functions in general.  
Our witness, \texttt{cvfn-1}, is a constant function and so
is clearly convex. This can be seen in Prog.~\ref{prog:cvfn-1}.
\begin{Program}
    \caption{Encapsulating a constant function \texttt{cvnf-1} and stating its convexity. \label{prog:cvfn-1}}
    \begin{lstlisting}
(encapsulate
 (((cvfn-1 *) => *)...)

 (local (defun cvfn-1 (x) (declare (ignore x)) 1337))
 ...
 (defthm cvfn-1-convex
  (implies (and (real-listp x) (real-listp y) (= (len y) (len x))
                (realp a) (<= 0 a) (<= a 1))
           (<= (cvfn-1 (vec-+ (scalar-* a x) (scalar-* (- 1 a) y)))
               (+ (* a (cvfn-1 x)) (* (- 1 a) (cvfn-1 y)))))) ...)
    \end{lstlisting}
\end{Program}

Explicitly, $a\cdot f$ is convex because 
\begin{equation}
af(\alpha x + (1-\alpha) y) \leq a( \alpha f(x) + (1-\alpha)f(y) ) = \alpha (af(x)) + (1-\alpha) \left( af(y) \right).
\end{equation}
In particular, we invoke convexity and need to distribute $a$. Moreover, this line of reasoning is not dependent on 
the definition of $f$ so we may disable the definition of \texttt{cvfn-1} in 
Prog.~\ref{prog:cvfn-1}.
This can be seen in Prog.~\ref{prog:af-convex}.

\begin{Program}
    \caption{Supressing the definition of \texttt{cvfn-1}, stating a special case of distributivity, and stating
    the convexity of $a\cdot f$.}\label{prog:af-convex}
    \begin{lstlisting}
(local (in-theory (disable (:d cvfn-1) (:e cvfn-1) (:t cvfn-1))))

(encapsulate ...
  ;;  factor out alpha
  (local (defthm lemma-1
   (implies (and (real-listp x) (real-listp y) (= (len y) (len x))
                 (realp a) (<= 0 a) (<= a 1)
                 (realp alpha) (<= 0 alpha))
            (= (+ (* a (* alpha (cvfn-1 x)))
                  (* (- 1 a) (* alpha (cvfn-1 y))))
               (* alpha
                  (+ (* a (cvfn-1 x))
                     (* (- 1 a) (cvfn-1 y))))))))

  (defthm a-*-cvfn-1-convex
   (implies (and (real-listp x) (real-listp y) (= (len y) (len x))
                 (realp a) (<= 0 a) (<= a 1)
                 (realp alpha) (<= 0 alpha))
            (<= (* alpha (cvfn-1 (vec-+ (scalar-* a x) (scalar-* (- 1 a) y))))
                (+ (* a (* alpha (cvfn-1 x)))
                   (* (- 1 a) (* alpha (cvfn-1 y))))))
   :hints (("GOAL" :in-theory (disable distributivity)
                   :use ((:instance cvfn-1-convex)
                         (:instance lemma-1))))))
    \end{lstlisting}
\end{Program}

We omit the formal proofs for the other claims of Thm.~\ref{thm:compose} 
as well as the proofs of convexity for the Euclidean norm $\|\cdot\|_2$ and  its square 
$\|\cdot\|_2^2$ and the inequality \footnote{
These proofs can be found in \texttt{convex.lisp}.
}
\begin{equation}
\alpha (1-\alpha)\|x - y\|^2 \leq \alpha \|x\|^2 + (1-\alpha) \|y\|^2.
\end{equation}

\section{Nesterov's Theorem in ACL2(r)}


\subsection{Approach}
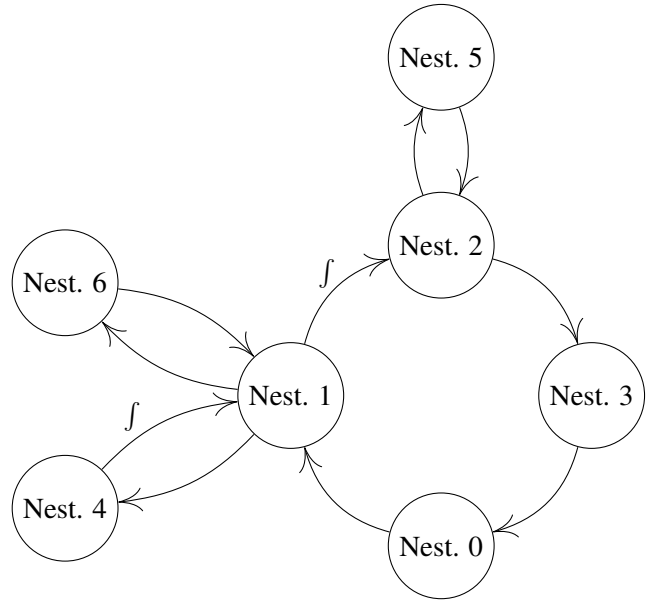
\begin{figure}
\begin{center}
\begin{tikzpicture}[
  dot/.style={draw,circle,inner sep=3pt}]
  \node[dot] (x) at (0,0) {Nest. 0};
  \node[dot] (1) at (-2,2) {Nest. 1};
  \node[dot] (2) at (0,4) {Nest. 2};
  \node[dot] (3) at (2,2) {Nest. 3};

\draw[-{>[scale = 3,length=3, width = 3]}]  (x) to[bend left = 30] (1);
\draw[-{>[scale = 3,length=3, width = 3]}]  (1) to[bend left = 30]node[pos=.4,above]{$\int$}(2);
\draw[-{>[scale = 3,length=3, width = 3]}]  (2) to[bend left = 30] (3);
\draw[-{>[scale = 3,length=3, width = 3]}]  (3) to[bend left = 30] (x);

\node[dot] (4) at (-5, .5) {Nest. 4};
\draw[-{>[scale = 3,length=3, width = 3]}]  (4) to[bend left = 20]node[pos=.25,above]{$\int$}(1);
\draw[-{>[scale = 3,length=3, width = 3]}]  (1) to[bend left = 20] (4) ;

\node[dot] (6) at (-5, 3.5) {Nest. 6};
\draw[-{>[scale = 3,length=3, width = 3]}]  (6) to[bend left = 20] (1);
\draw[-{>[scale = 3,length=3, width = 3]}]  (1) to[bend left = 20] (6);

\node[dot] (5) at (0, 6.5) {Nest. 5};
\draw[-{>[scale = 3,length=3, width = 3]}]  (5) to[bend left = 20] (2);
\draw[-{>[scale = 3,length=3, width = 3]}]  (2) to[bend left = 20] (5);

\end{tikzpicture}
\caption{Nesterov's proof of Thm.~\ref{thm:nesterov}. Here Nest. 0 is Lipschitz continuity.
Integration is denoted by $\int$.
}
\label{fig:nestapproach}
\end{center}
\end{figure}

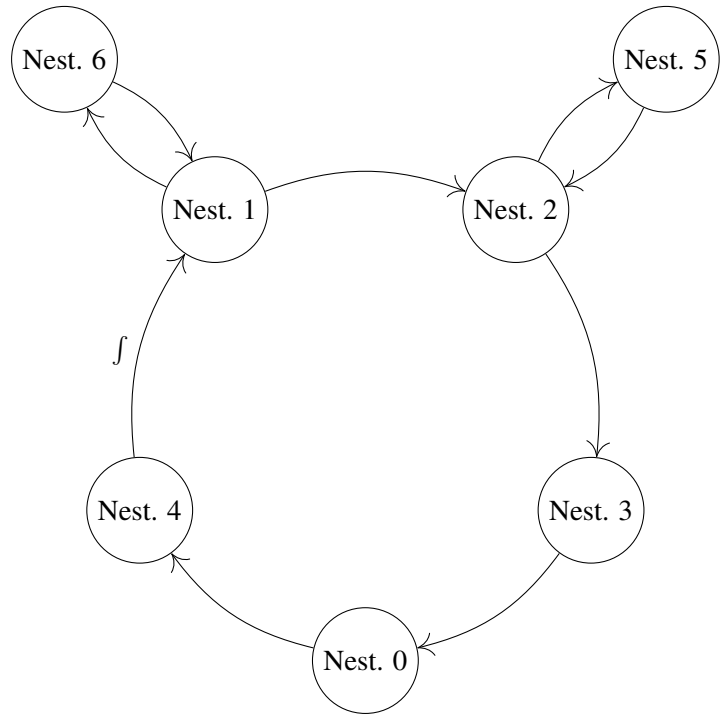
\begin{figure}
\begin{center}
\begin{tikzpicture}[
  dot/.style={draw,circle,inner sep=3pt}]
  \node[dot] (x) at (0,0) {Nest. 0};
  \node[dot] (4) at (-3,2) {Nest. 4};
  \node[dot] (1) at (-2,6) {Nest. 1};
  \node[dot] (2) at (2,6) {Nest. 2};
  \node[dot] (3) at (3,2) {Nest. 3};

\draw[-{>[scale = 3,length=2, width = 3]}]  (x) to[bend left = 20] (4);
\draw[-{>[scale = 3,length=2, width = 3]}]  (4) to[bend left = 20] node[midway,left]{$\int$}(1);
\draw[-{>[scale = 3,length=2, width = 3]}]  (1) to[bend left = 20] (2);
\draw[-{>[scale = 3,length=2, width = 3]}]  (2) to[bend left = 20] (3);
\draw[-{>[scale = 3,length=2, width = 3]}]  (3) to[bend left = 20] (x);

  \node[dot] (6) at (-4,8) {Nest. 6};
\draw[-{>[scale = 3,length=2, width = 3]}]  (1) to[bend left = 20] (6);
\draw[-{>[scale = 3,length=2, width = 3]}]  (6) to[bend left = 20] (1);

  \node[dot] (5) at (4,8) {Nest. 5};
\draw[-{>[scale = 3,length=2, width = 3]}]  (2) to[bend left = 20] (5);
\draw[-{>[scale = 3,length=2, width = 3]}]  (5) to[bend left = 20] (2);
\end{tikzpicture}
\caption{Another proof of Thm.~\ref{thm:nesterov}. Here Nest. 0 is Lipschitz continuity. Integration is denoted by $\int$.}
\label{fig:altapproach}
\end{center}
\end{figure}

In \cite{Nesterov2004}, Nesterov provided a proof that followed the
structure visualised by Fig.~\ref{fig:nestapproach}.
Nesterov's proof, however, uses techniques that are not amenable to proofs in ACL2(r).
In particular, integration is used multiple times to show some inequalities.
However, integration in ACL2(r) is dependent 
on the function that is being integrated~\cite{Kaufmann2000}. This places 
extra obligations on the user. The alternate approach shown in Fig.~\ref{fig:altapproach}
requires fewer instances of integration than Fig.~\ref{fig:nestapproach}. Moreover, 
Fig.~\ref{fig:altapproach} has fewer implications to prove in general.
The primary difference in our approach is that we prove \ref{nest:4} from a 
straightforward application of Cauchy-Schwarz and omit 
\ref{nest:1} implies \ref{nest:4}.

Stating a theorem about functions in ACL2 is an unnatural endeavour because ACL2 is a 
theorem prover for first-order logic so we cannot predicate over sets in general.
The natural solution is
 to leverage encapsulation and functional instantiation 
to obtain pseudo-higher order behaviour.
However, this means that the desired function for which the user wishes to apply 
Thm.~\ref{thm:nesterov} must pass the theorems within the encapsulation. 
To formalise the theorem in its greatest generality, it is necessary to suppress the 
definition of the witness function in the encapsulation 
and instead prove the theorems based on the properties of the function.
To use functional instantiation, these properties must be proven for the desired function.
Thus, we aim to minimize the number of properties that the user must show, and derive
as much as possible within the encapsulation.
In our case, the user obligations are the theorems and identities involving the
continuity, derivative, and
integral of the encapsulated functions as well as any forms explicitly involving 
the dimension of the space.

The encapsulated functions include the multivariate function of interest \texttt{mvfn} and
its derivative \texttt{nabla-mvfn},
the function that evaluates to the Lipschitz constant \texttt{L}, 
a helper function \texttt{phi} based on \texttt{mvfn} and its derivative \texttt{nabla-phi}, 
the recognizer for vectors with standard real entries 
\texttt{standard-vecp}, and the function, \texttt{DIM},
 that evaluates to the dimension $n$
of the vector space.

The fundamental 
challenges of formalising $\Rn{n}$ are explored in \cite{Kwan2018-cs}. 
In the context of this paper, however, the particular difficulties to note are 
the recognizer for vectors with standard entries, exhibiting the relationship between 
vectors with infinitesimal entries and the norm, 
and
invoking two copies of certain inequalities. More details will be discussed as we proceed.

\subsection{Basic Definitions \& Lemmas}\label{sec:basic}

As an example of the formalisation, the definition of continuity for 
multivariate functions can be seen in Prog.~\ref{prog:acl2-cont}.
\begin{Program}
    \caption{Continuity for a multivariate function \texttt{mvfn}.
\label{prog:acl2-cont}}
\begin{lstlisting}
 (defthm mvfn-is-continuous-3
  (implies (and (real-listp vec1) (real-listp vec2)
                (= (len vec2) (len vec1)) (= (len vec1) (DIM))
                (i-small (eu-metric vec1 vec2)))
           (i-small (abs (- (mvfn vec1) (mvfn vec2))))))
\end{lstlisting}
\end{Program}
The hypotheses ensure the entries are both real vectors of the same dimension 
and \texttt{i-small} is the recognizer for infinitesimals. 
This form is consistent with 
the non-standard analysis definition of continuity. 


The prohibition of non-classical recursive functions and the necessity of 
a recognizer for vectors with standard entries forces the recognizer to be defined
with a specific $n$
within the encapsulation.
An encapsulation requires nevertheless a witness for the convex function of interest
(which in this case happens to be the function $f(x) = 42$) so we set $n=2$. 
The recognizer \texttt{standard-vecp} simply checks whether a 
vector of dimension 2 has standards in both its 
coordinates.\footnote{The rest of the basic definitions and theorems can be found in \texttt{nesterov-1.lisp}.}

Another recurring and particularly useful theorem is the Cauchy-Schwarz inequality. 
It was formalised in \cite{Kwan2018-cs} and has applications in proving the 
triangle inequality and the properties of $\Rn{n}$ as a metric space.
We use a specific version of the inequality that can be seen in Prog.~\ref{prog:cs2}.
\begin{Program}
    \caption{A version of the Cauchy-Schwarz inequality used in 
        our proof of Thm. \ref{thm:nesterov}. \label{prog:cs2} }
        \begin{lstlisting}
(defthm cauchy-schwarz-2
 (implies (and (real-listp u) (real-listp v) (= (len u) (len v)))
          (<= (abs (dot u v))
              (* (eu-norm u) (eu-norm v))))...)
\end{lstlisting}
\end{Program}

There are several proofs of implications in Fig.~\ref{fig:altapproach} that follow almost
immediately from the mentioned definitions and lemmas. We outline one of them:
Nest. 0 implies \ref{nest:4}.
The proof mimics  the chain of inequalities
\begin{equation}\label{eq:ineq-0-implies-ineq-4}
L\|x-y\|^2\geq \|f^\prime(x)-f^\prime(y)\|\cdot
\|x-y\|\geq\langle f^\prime(x)-f^\prime(y),x-y\rangle
\end{equation}
where the first inequality follows from Lipschitz continuity
and the second inequality follows from Cauchy-Schwarz.
We begin with the first inequality of Eq.~\ref{eq:ineq-0-implies-ineq-4} in Prog.~\ref{prog:lem-2}.
\begin{Program}
    \caption{The first inequality follows from Lipschitz continuity. \label{prog:lem-2}}
    \begin{lstlisting}
;; ||f'(x) - f'(y)|| <= L||x - y|| implies
;; ||f'(x) - f'(y)|| ||x - y|| <= L||x - y||^2
(local (defthm lemma-2
 (implies (and (real-listp x) (real-listp y) 
               (= (len y) (len x)) (= (len x) (DIM))
               (<= (eu-norm (vec-- (nabla-mvfn x) (nabla-mvfn y))) 
                   (* (L) (eu-norm (vec-- x y)))))
          (<= (* (eu-norm (vec-- (nabla-mvfn x) (nabla-mvfn y))) 
                 (eu-norm (vec-- x y)))
              (* (L) (eu-norm (vec-- x y)) 
                     (eu-norm (vec-- x y)))))...))
\end{lstlisting}
\end{Program}
Applying the Cauchy-Schwarz inequality from Prog.~\ref{prog:cs2} gives the second inequality of
Eq.~\ref{eq:ineq-0-implies-ineq-4} under 
absolute values as seen in Prog.~\ref{prog:lem-3}.
\begin{Program}
        \caption{The second inequality follows from Cauchy-Schwarz. \label{prog:lem-3}}
        \begin{lstlisting}
;; |<f'(x) - f'(y), x - y>| <= L||x - y||^2
(local (defthm lemma-3
 (implies (and (real-listp x) (real-listp y) 
               (= (len y) (len x)) (= (len x) (DIM))
               (<= (eu-norm (vec-- (nabla-mvfn x) (nabla-mvfn y))) 
                   (* (L) (eu-norm (vec-- x y)))))
         (<= (abs (dot (vec-- (nabla-mvfn x) (nabla-mvfn y)) 
                       (vec-- x y)))
             (* (L) (eu-norm (vec-- x y)) 
                    (eu-norm (vec-- x y)))))
 :hints (("GOAL" :use ((:instance lemma-2)
                       ...
                       (:instance cauchy-schwarz-2
                                  (u (vec-- (nabla-mvfn x) 
                                            (nabla-mvfn y)))
                                  (v (vec-- x y))))))))
\end{lstlisting}
\end{Program}
Eliminating absolute values gives the desired inequality (in an ``expanded" form) as seen in Prog.~\ref{prog:0-implies-4-expanded}.
\begin{Program}
    \caption{The desired implication in an expanded form. \label{prog:0-implies-4-expanded}}
    \begin{lstlisting}
;; <f'(x) - f'(y), x - y> <= L ||x - y||^2
(defthm ineq-0-implies-ineq-4-expanded
 (implies (and (real-listp x) (real-listp y) 
               (= (len y) (len x)) (= (len x) (DIM))
               (<= (eu-metric (nabla-mvfn x) (nabla-mvfn y))
                   (* (L) (eu-metric x y))))
         (<= (dot (vec-- (nabla-mvfn x) (nabla-mvfn y)) (vec-- x y))
             (* (L) (metric^2 x y))))
 :hints (("GOAL" :use (...(:instance lemma-3)))))
\end{lstlisting}
\end{Program}
To state the implication in its full generality and for reasons to appear in the next section,
we use Skolem functions to replace the inequalities in the theorem.
The function definitions can be seen in Prog.~\ref{prog:sk}.
The theorem then becomes of the form seen in Prog.~\ref{prog:ineq-0-implies-ineq-4}
where the hypotheses ensure that we are dealing with real vectors.
All the implications, whether they follow straight from the definitions or otherwise,
are of this form.\footnote{The rest of the
Skolem functions can be found in \texttt{nesterov-4.lisp}}

The other implications that follow mainly from the definitions are 
\ref{nest:4} implies \ref{nest:1} and  \ref{nest:1} implies \ref{nest:2}.

\begin{Program}
    \caption{Skolem function definitions that allow us to invoke
        the \texttt{forall} quantifier.}\label{prog:sk}
        \begin{lstlisting}
;; Lipschitz continuity ||f'(x) - f'(y)|| <= L ||x - y||
(defun-sk ineq-0 (L)
 (forall (x y)
  (<= (eu-metric (nabla-mvfn x) (nabla-mvfn y))
      (* L (eu-metric x y))))...)
...
;; <f'(x) - f'(y), x - y> <= L ||x - y||^2
(defun-sk ineq-4 (L)
 (forall (x y)
  (<= (dot (vec-- (nabla-mvfn x) (nabla-mvfn y)) (vec-- x y))
      (* L (metric^2 x y))))...)
\end{lstlisting}
\end{Program}

\begin{Program}
    \caption{The desired implication. \label{prog:ineq-0-implies-ineq-4}}
    \begin{lstlisting}
(defthm ineq-0-implies-ineq-4
  (implies (and (hypotheses (ineq-4-witness (L)) (DIM))
                (ineq-0 (L)))
           (ineq-4 (L)))...)
\end{lstlisting}
\end{Program}

\subsection{Challenging Issues}
Here we outline some of the challenges we encountered during our formalisation 
of Thm.~\ref{thm:nesterov}.
Several of these issues involve the proofs of the remaining lemmas,
which
all require some user intervention
beyond simple algebraic manipulation.
Here we discuss two such instances.
The others are omitted because we solve them similarly.
Finally, we discuss the final form of Thm.~\ref{thm:nesterov} and the various considerations regarding 
it and alternative approaches.

\subsubsection{Instantiating Inequalities}\label{sec:ineq}
The proof of \ref{nest:2} implies \ref{nest:3} 
amounts to
adding two copies of \ref{nest:2} with $x$, $y$ swapped.
This induces issues regarding the proof of the implication. 
The natural form of the lemma would involve \ref{nest:2} 
among the hypotheses as in Prog.~\ref{prog:natural}.

\begin{Program}
    \caption{An ``obvious" way to state \ref{nest:2}  implies \ref{nest:3}.}\label{prog:natural}
    \begin{lstlisting}
(defthm ineq-2-implies-ineq-3
  (implies (and (real-listp x) (real-listp y) 
                (= (len y) (len x)) (= (len x) (DIM))
                (ineq-2 (L)))
           (ineq-3 (L)))...)
\end{lstlisting}
\end{Program}

However, to instantiate a copy of \ref{nest:2} with swapped $x$, $y$ in such a form would be 
equivalent to
\begin{equation}
\forall x,y, (P(x,y)\implies P(y,x))
\end{equation}
where $P$ is a predicate (in this case equivalent to \ref{nest:2}),
 which is not necessarily true.
The form we wish to have is 
\begin{equation}
(\forall x,y, P(x,y))\implies (\forall x,y, P(y,x)).
\end{equation}
In order to instantiate another copy of \ref{nest:2} within the implication requires 
quantifiers within the theorem statement. The usual approach 
involves using Skolem functions to introduce quantified variables.
We can now instantiate the two copies with swapped variables as in Prog.~\ref{prog:swap}.
\begin{Program}
    \caption{Instantiating two copies of \ref{nest:2} with swapped
        variables.}\label{prog:swap}
\begin{lstlisting}
(defthm ineq-2-expanded-v1
  (implies (ineq-2 (L))
           (and (<= (+ (mvfn x) 
                       (dot (nabla-mvfn x) (vec-- y x))
                       (* (/ (* 2 (L))) 
                          (metric^2 (nabla-mvfn x) (nabla-mvfn y))))
                    (mvfn y))
                (<= (+ (mvfn y) 
                       (dot (nabla-mvfn y) (vec-- x y))
                       (* (/ (* 2 (L))) 
                          (metric^2 (nabla-mvfn y) (nabla-mvfn x))))
                    (mvfn x))))...)
\end{lstlisting}
\end{Program}

We also considered simply including two copies of the inequality with swapped variables 
among the hypotheses. 
This has two advantages. 
Firstly, with such a form, 
the lemma becomes stronger because the hypothesis
$P(x,y)\land P(y,x)$ is weaker than $\forall x,y, P(x,y)$. 
Secondly, the lemma is slightly easier
to pass in ACL2(r).
However, the primary drawback is that this form is inconsistent with the 
other lemmas and
 the final form of Nesterov's theorem becomes less 
elegant
(eg. showing \ref{nest:1} implies \ref{nest:2} would also require two copies of \ref{nest:1}).

The lemmas \ref{nest:1} implies \ref{nest:6} and \ref{nest:2} implies \ref{nest:5} also requires instantiating 
multiple copies of the antecedent inequalities (albeit with different vectors).

\subsubsection{Taking Limits}
In the language of non-standard analysis, limits amount to taking standard-parts.
For example, $\lim_{x\to a}f(x) $ is equivalent to $\st (f(x))$ when $x-a$ is an infinitesimal.
However, for products, say, $xy$, the identity $\st(xy)=\st(x)\st(y)$ only holds when 
$x$, $y$ are both finite reals. 
In the proof of \ref{nest:6} implies \ref{nest:1}, there is a step that requires taking the limit
of 
$(1-\alpha)\|y-x\|^2$ as $\alpha\to 0$. Now, if $\alpha>0$ is an infinitesimal,
\begin{equation}
\st((1-\alpha)\|y-x\|^2) = \st(1-\alpha)\st(\|y-x\|^2)=\|y-x\|^2
\end{equation}
is easy to satisfy when $x$, $y$ are vectors with standard  real
components.
Moreover, requiring variables to be standard is consistent with some
instances of single variable theorems (eg. the product of continuous functions is continuous).
It then remains to state such a hypothesis using, say, a recognizer \texttt{standard-vecp} as 
in Prog.~\ref{prog:6-implies-1-expanded}.
\begin{Program}
    \caption{Introducing \texttt{standard-vecp} into the hypotheses.}\label{prog:6-implies-1-expanded}
    \begin{lstlisting}
(defthm ineq-6-implies-ineq-1-expanded 
 (implies (and (real-listp x) (real-listp y) 
               (= (len y) (len x)) (= (len x) (DIM))
               (realp alpha) (i-small alpha)
               (< 0 alpha) (<= alpha 1) 
               (standard-vecp x) (standard-vecp y)
               (<= (+ (* alpha (mvfn y))
                      (* (- 1 alpha) (mvfn x)))
                   (+ (mvfn (vec-+ (scalar-* alpha y) 
                                   (scalar-* (- 1 alpha) x)))
                      (* (/ (L) 2) alpha (- 1 alpha) (metric^2 y x)))))
          (<= (mvfn y)
              (+ (mvfn x)
                 (dot (nabla-mvfn x) (vec-- y x))
                 (* (/ (L) 2) (metric^2 y x)))))...)
\end{lstlisting}
\end{Program}
The natural approach to defining \texttt{standard-vecp} would be to simply
recurse on the length of a vector applying \texttt{standardp} to each entry. 
However, \texttt{standardp} is non-classical
and this definition encounters a common issue throughout our ACL2(r) formalisation in that
it is a non-classical recursive function. We discuss the subtleties of this problem 
more in~\cite{Kwan2018-cs}. Because \texttt{standard-vecp} is dependent on the dimension,
our solution is to encapsulate the function and prove the necessary theorems involving it
(eg. \texttt{metric\textasciicircum2} is \texttt{standardp} on \texttt{standard-vecp} values). 
For the 
case $n=2$, we simply check the length of the vector is two and that both entries are 
standard reals.
A final note regarding the lemma is the hypothesis $\alpha>0$ replacing the weaker
$\alpha\geq0$ since part of the proof depends on dividing by $\alpha$.

The other lemma that requires similar hypotheses is the implication \ref{nest:5} implies \ref{nest:2}.

\subsubsection{Final Form of Nesterov's Theorem}
Finally, we discuss our final form of 
Thm.~\ref{thm:nesterov} as well as several alternatives
and their considerations.  The final form can be seen in Prog.~\ref{prog:nesterov}.
The function \texttt{hypotheses} ensure that we are dealing with 
real vectors of the correct dimension. The function \texttt{st-hypotheses}
ensure that the vectors have standard entries due to the necessity of taking 
limits. 
The function \texttt{alpha-hypotheses} is the same as \texttt{hypotheses}
but includes the hypothesis that $\alpha\in[0,1]$.
The function \texttt{alpha->-0-hypotheses} ensures that $\alpha>0$ for the case 
of taking limits.

\begin{Program}
    \caption{The final statement of Thm.~\ref{thm:nesterov}. \label{prog:nesterov}}
    \begin{lstlisting}
(defthm nesterov
 ;; theorem statement
 (implies (and (hypotheses (ineq-0-witness (L)) (DIM))
	            (hypotheses (ineq-1-witness (L)) (DIM))
	            (hypotheses (ineq-2-witness (L)) (DIM))
	            (hypotheses (ineq-3-witness (L)) (DIM))
	            (hypotheses (ineq-4-witness (L)) (DIM))
	            (st-hypotheses (ineq-1-witness (L)))
	            (st-hypotheses (ineq-2-witness (L)))
	            (alpha-hypotheses (ineq-5-witness (L)) (DIM))
	            (alpha-hypotheses (ineq-6-witness (L)) (DIM))
	            (alpha->-0-hypotheses (ineq-5 (L)))
	            (alpha->-0-hypotheses (ineq-6 (L)))
	            (or (ineq-0 (L)) (ineq-1 (L)) (ineq-2 (L)) (ineq-3 (L))
                   	(ineq-4 (L)) (ineq-5 (L)) (ineq-6 (L))))
	  (and (ineq-0 (L)) (ineq-1 (L)) (ineq-2 (L)) (ineq-3 (L))
               (ineq-4 (L)) (ineq-5 (L)) (ineq-6 (L))))
 ;; hints, etc. for ACL2(r)
 ...)
\end{lstlisting}
\end{Program}

In Sec.~\ref{sec:basic}, we already mentioned, for each lemma,
the basic structure involving Skolem functions. 
In Sec.~\ref{sec:ineq}, we cited elegance and ease of instantiation as 
reasons for using Skolem functions.
Because stating even the shorter inequalities in Polish notation
 would quickly become 
awkward and unintelligible (eg. Prog.~\ref{prog:ineq-5-polish}),
it became desirable for us to define the inequalities in a clean and clear manner.
One simple approach would be to define the inequalities as ACL2 functions or macros. 
Unfortunately, during the course of our formalisation, 
we found that the rewriter would
be tempted to ``simplify" or otherwise change the form of the 
inequality via arithmetic rules.
This made applications of certain theorems more involved and arduous than necessary. 
Therefore, it would be necessary to disable the function definitions anyways.
In addition to permitting instantiations of inequalities with 
different vectors within a single theorem statement,
Skolem functions would allow us to suppress or ``hide" the explicit 
inequality
thus providing a clear, concise, and compact package.

\begin{Program}
    \caption{\ref{nest:5} in Polish notation. \label{prog:ineq-5-polish}}
    \begin{lstlisting}
(<= (+ (mvfn (vec-+ (scalar-* alpha y) (scalar-* (- 1 alpha) x)))
       (* (/ (* 2 (L))) (* alpha (- 1 alpha) (metric^2 (nabla-mvfn y) 
                                         	       (nabla-mvfn x)))))
    (+ (* alpha (mvfn y)) (* (- 1 alpha) (mvfn x))))
\end{lstlisting}
\end{Program}

On the other hand, 
this form has the unfortunate drawback of making the proof of the theorem 
slightly more involved. By introducing Skolem functions we also introduce 
the necessity of witness functions;
proving the lemmas in terms of 
the witness functions may occasionally become onerous.
For example, to state the hypotheses that the entries of the
 witness functions are real vectors of the same dimension, we 
would like to define a \texttt{hypotheses} function.
 However,  explicitly exhibiting the
witness functions within the definition of \texttt{hypotheses} leads to a signature 
mismatch. 
We instead 
pass the witness function to \texttt{hypotheses} as an argument.

%


\section{Conclusion}

In this paper, we presented a set of theorems for reasoning about convex functions in 
ACL2(r). We also discussed some of the challenges of formalising 
a proof that relies heavily on informally well-established and intuitive notions. 
Examples include 
translating statements in classical multivariate analysis into non-standard analysis and
instantiating quantified statements using Skolem functions.
Our particular interest in this work are the potential applications to verifying,
among other areas,
optimization algorithms used in machine learning.
To this end, we chose a theorem of Nesterov's to serve as an example of the 
analytical reasoning possible in our formalisation. 
The natural next step would be to develop a proper 
theory of multivariate calculus to further automate the reasoning
about optimisation algorithms.

%

\bibliography{refs}

\begin{thebibliography}{10}
\providecommand{\bibitemdeclare}[2]{}
\providecommand{\surnamestart}{}
\providecommand{\surnameend}{}
\providecommand{\urlprefix}{Available at }
\providecommand{\url}[1]{\texttt{#1}}
\providecommand{\href}[2]{\texttt{#2}}
\providecommand{\urlalt}[2]{\href{#1}{#2}}
\providecommand{\doi}[1]{doi:\urlalt{http://dx.doi.org/#1}{#1}}
\providecommand{\bibinfo}[2]{#2}

\bibitemdeclare{book}{cutland}
\bibitem{cutland}
\bibinfo{author}{Leif~O. \surnamestart Arkeryd\surnameend},
  \bibinfo{author}{Nigel~J. \surnamestart Cutland\surnameend} \&
  \bibinfo{author}{C.~Ward~Henson \surnamestart (Eds.)\surnameend}
  (\bibinfo{year}{1997}): \emph{\bibinfo{title}{Nonstandard Analysis: Theory
  and Applications}}, \bibinfo{edition}{1st} edition.
\newblock {\sl \bibinfo{series}{Nato Science Series C: 493}}~,
  \bibinfo{publisher}{Springer Netherlands}, \doi{10.1007/978-94-011-5544-1}.

\bibitemdeclare{book}{boyd}
\bibitem{boyd}
\bibinfo{author}{Stephen \surnamestart Boyd\surnameend} \&
  \bibinfo{author}{Lieven \surnamestart Vandenberghe\surnameend}
  (\bibinfo{year}{2004}): \emph{\bibinfo{title}{Convex Optimization}}.
\newblock \bibinfo{publisher}{Cambridge University Press},
  \doi{10.1017/CBO9780511804441}.

\bibitemdeclare{incollection}{Gamboa2000}
\bibitem{Gamboa2000}
\bibinfo{author}{Ruben \surnamestart Gamboa\surnameend} (\bibinfo{year}{2000}):
  \emph{\bibinfo{title}{Continuity and Differentiability}}.
\newblock In \bibinfo{editor}{Matt \surnamestart Kaufmann\surnameend},
  \bibinfo{editor}{Panagiotis \surnamestart Manolios\surnameend} \&
  \bibinfo{editor}{J.~Strother \surnamestart Moore\surnameend}, editors: {\sl
  \bibinfo{booktitle}{Computer-Aided Reasoning: ACL2 Case Studies}},
  \bibinfo{publisher}{Springer US}, \bibinfo{address}{Boston, MA}, pp.
  \bibinfo{pages}{301--315}, \doi{10.1007/978-1-4757-3188-0_18}.

\bibitemdeclare{article}{Gamboa2001}
\bibitem{Gamboa2001}
\bibinfo{author}{Ruben~A. \surnamestart Gamboa\surnameend} \&
  \bibinfo{author}{Matt \surnamestart Kaufmann\surnameend}
  (\bibinfo{year}{2001}): \emph{\bibinfo{title}{Nonstandard Analysis in ACL2}}.
\newblock {\sl \bibinfo{journal}{Journal of Automated Reasoning}}
  \bibinfo{volume}{27}(\bibinfo{number}{4}), pp. \bibinfo{pages}{323--351},
  \doi{10.1023/A:1011908113514}.

\bibitemdeclare{inproceedings}{harrison2007}
\bibitem{harrison2007}
\bibinfo{author}{John \surnamestart Harrison\surnameend}
  (\bibinfo{year}{2007}): \emph{\bibinfo{title}{Formalizing Basic Complex
  Analysis}}.
\newblock In \bibinfo{editor}{R.~\surnamestart Matuszewski\surnameend} \&
  \bibinfo{editor}{A.~\surnamestart Zalewska\surnameend}, editors: {\sl
  \bibinfo{booktitle}{From Insight to Proof: Festschrift in Honour of Andrzej
  Trybulec}}, {\sl \bibinfo{series}{Studies in Logic, Grammar and Rhetoric}}
  \bibinfo{volume}{10(23)}, \bibinfo{publisher}{University of Bia{\l}ystok},
  pp. \bibinfo{pages}{151--165}.
\newblock \urlprefix\url{http://mizar.org/trybulec65/}.

\bibitemdeclare{book}{jacob}
\bibitem{jacob}
\bibinfo{author}{Nathan \surnamestart Jacobson\surnameend}
  (\bibinfo{year}{1985}): \emph{\bibinfo{title}{Basic Algebra I}},
  \bibinfo{edition}{2nd} edition.
\newblock \bibinfo{publisher}{Dover Publications}.

\bibitemdeclare{incollection}{Kaufmann2000}
\bibitem{Kaufmann2000}
\bibinfo{author}{Matt \surnamestart Kaufmann\surnameend}
  (\bibinfo{year}{2000}): \emph{\bibinfo{title}{Modular Proof: The Fundamental
  Theorem of Calculus}}.
\newblock In \bibinfo{editor}{Matt \surnamestart Kaufmann\surnameend},
  \bibinfo{editor}{Panagiotis \surnamestart Manolios\surnameend} \&
  \bibinfo{editor}{J.~Strother \surnamestart Moore\surnameend}, editors: {\sl
  \bibinfo{booktitle}{Computer-Aided Reasoning: ACL2 Case Studies}},
  \bibinfo{publisher}{Springer US}, \bibinfo{address}{Boston, MA}, pp.
  \bibinfo{pages}{75--91}, \doi{10.1007/978-1-4757-3188-0_6}.

\bibitemdeclare{conference}{Kwan2018-cs}
\bibitem{Kwan2018-cs}
\bibinfo{author}{Carl \surnamestart Kwan\surnameend} \&
  \bibinfo{author}{Mark~R. \surnamestart Greenstreet\surnameend}
  (\bibinfo{year}{2018}): \emph{\bibinfo{title}{Real Vector Spaces and the
  Cauchy-Schwarz Inequality in ACL2(r)}}.
\newblock In: {\sl \bibinfo{booktitle}{{\rm Proceedings 15th International
  Workshop on the} ACL2 Theorem Prover and its Applications, {\rm Austin,
  Texas, USA, November 5-6, 2018}}}, \bibinfo{series}{\thisvolume{7}}.

\bibitemdeclare{book}{lang}
\bibitem{lang}
\bibinfo{author}{Serge \surnamestart Lang\surnameend} (\bibinfo{year}{2002}):
  \emph{\bibinfo{title}{Algebra}}, \bibinfo{edition}{3rd} edition.
\newblock {\sl \bibinfo{series}{Graduate Texts in Mathematics 211}}~,
  \bibinfo{publisher}{Springer-Verlag New York},
  \doi{10.1007/978-1-4613-0041-0}.

\bibitemdeclare{book}{loeb}
\bibitem{loeb}
\bibinfo{author}{Peter~A. \surnamestart Loeb\surnameend} \&
  \bibinfo{author}{Manfred P.~H. \surnamestart Wolff\surnameend}
  (\bibinfo{year}{2015}): \emph{\bibinfo{title}{Nonstandard Analysis for the
  Working Mathematician}}, \bibinfo{edition}{2nd} edition.
\newblock \bibinfo{publisher}{Springer Netherlands},
  \doi{10.1007/978-94-017-7327-0}.

\bibitemdeclare{book}{Nesterov2004}
\bibitem{Nesterov2004}
\bibinfo{author}{Yurii \surnamestart Nesterov\surnameend}
  (\bibinfo{year}{2004}): \emph{\bibinfo{title}{Introductory Lectures on Convex
  Optimization}}, \bibinfo{edition}{1st} edition.
\newblock {\sl \bibinfo{series}{Applied Optimization 87}}~,
  \bibinfo{publisher}{Springer US}, \doi{10.1007/978-1-4419-8853-9}.

\bibitemdeclare{book}{robinson}
\bibitem{robinson}
\bibinfo{author}{Abraham \surnamestart Robinson\surnameend}
  (\bibinfo{year}{1966}): \emph{\bibinfo{title}{Non-Standard Analysis}}.
\newblock \bibinfo{publisher}{North-Holland Publishing Company}.

\bibitemdeclare{book}{roman}
\bibitem{roman}
\bibinfo{author}{Steven \surnamestart Roman\surnameend} (\bibinfo{year}{2008}):
  \emph{\bibinfo{title}{Advanced Linear Algebra}}, \bibinfo{edition}{3rd}
  edition.
\newblock {\sl \bibinfo{series}{Graduate Texts in Mathematics 135}}~,
  \bibinfo{publisher}{Springer-Verlag New York},
  \doi{10.1007/978-0-387-72831-5}.

\bibitemdeclare{book}{babyrudin}
\bibitem{babyrudin}
\bibinfo{author}{Walter \surnamestart Rudin\surnameend} (\bibinfo{year}{1976}):
  \emph{\bibinfo{title}{Principles of Mathematical Analysis}},
  \bibinfo{edition}{3rd} edition.
\newblock {\sl \bibinfo{series}{International Series in Pure and Applied
  Mathematics}}~, \bibinfo{publisher}{McGraw-Hill}.

\bibitemdeclare{book}{shilov}
\bibitem{shilov}
\bibinfo{author}{Georgi~E. \surnamestart Shilov\surnameend}
  (\bibinfo{year}{1977}): \emph{\bibinfo{title}{Linear Algebra}}.
\newblock \bibinfo{publisher}{Dover Publications}.

\end{thebibliography}
\bibliographystyle{eptcs}

\end{document}